\RequirePackage[colorlinks,allcolors=cyan]{hyperref}
\documentclass[draft]{agujournal2019}
\usepackage{url}
\usepackage{doi}

\usepackage{lineno}
\usepackage{graphicx}
\usepackage[final]{microtype}
\usepackage{amsmath}
\usepackage{amsfonts}
\usepackage{amssymb}
\usepackage{booktabs}
\let\linenumbers\relax

\journalname{Journal of Geophysical Research: Machine Learning and Computation}

\begin{document}

\title{Expert lithological labels do not improve learning when used to
predict marine sediment properties}

\authors{John M. Aiken\affil{1}, William Gilpin\affil{2},
Thorsten W. Becker\affil{3,4}}

\affiliation{1}{University of Oslo, Oslo, Norway}
\affiliation{2}{Department of Physics, University of Texas at Austin,
Austin, TX, USA}
\affiliation{3}{Institute for Geophysics, Jackson School of Geosciences,
University of Texas at Austin, Austin, TX, USA}
\affiliation{4}{Department of Earth and Planetary Sciences, Jackson School
of Geosciences, University of Texas at Austin, Austin, TX, USA}

\correspondingauthor{John M. Aiken}{johnm.aiken@gmail.com}

\begin{keypoints}
\item Six routinely logged core measurements recover geologist-assigned lithology without the need for expert labels.
\item Adding lithology labels to training does not improve any metric, and at high weight degrades reconstruction of the measurements
\item Depth alone is not predictive of unseen boreholes, and the
embedding beats depth on eight of nine held-out properties
\end{keypoints}

\begin{abstract}
Scientific drilling has produced large archives of physical, chemical
and magnetic measurements linked to descriptions of the rock recovered.
These archives are mostly incomplete: instruments sample at different
intervals, recovery is partial, and destructive analyses preclude later
measurement of the same material. We train an self-supervised embedding
model on the LILY ocean drilling database using six routinely logged
properties (bulk density, magnetic susceptibility, natural gamma
radiation, and red, green and blue reflectance) across 236,963
observations from 296 boreholes, with a masked variational autoencoder
that has a heteroscedastic Gaussian decoder and learnable mask tokens.
The embeddings separate geologist-assigned lithologies as well as a
variant trained with those labels, and increasing the weight on the
labels degrades reconstruction while leaving classification unchanged.
The labels are recoverable from the measurements geologists use to
assign them, so training on them adds nothing. Lithological categories
compress continuous measurements into classes, and compressing them
discards what the measurements record about the processes that produced
the rock. The embeddings also predict physical properties that were
absent from training, on boreholes the model has never seen. While depth
should account for this since sediments compact with burial, it does
not. On held-out boreholes depth alone reaches a median $R^2$ of
$-0.02$, against 0.20 for the embedding.
\end{abstract}

\section*{Plain Language Summary}
Scientific ships and rigs drill into the sea floor and bring up
cylinders of rock and mud. Machines scan each cylinder to record its
density, magnetization, radioactivity, and color. Geologists also look
at the core and write down what kind of rock it is. We trained a
computer model on the machine measurements alone, without showing it any
of the geologists' names for the rocks. The model learned a compact
numerical summary of each piece of core. From that summary alone we
could tell the rock types apart about as well as we could using the
geologists' own labels. When we forced the model to pay attention to the
labels during training, nothing improved, and the model became worse at
describing the measurements. The reason is that geologists name rocks by
looking at color, density and texture, so the names are already
contained in the measurements. The same compact summary also let us
estimate properties that were never measured on most of the core. This
suggests that machine measurements collected as a matter of routine
carry more information about the sub-surface than they are usually
credited with.

\section{Introduction}
\label{sec:intro}

The Earth's sub-surface is sampled through boreholes, which give a
variety of dense measurements along depth, including physical, chemical
and sometimes biological analyses. Borehole data come from ocean drilling
programs, from continental programs such as the International
Continental Scientific Drilling Program (ICDP), and from mineral,
geothermal and hydrocarbon exploration. This paper explores the impact
that lithology labeling of marine sediment cores has in the context of
these varied measurements, and its impact on prediction.

The International Ocean Discovery Program (IODP) and its predecessors
(the Deep Sea Drilling Project and the Ocean Drilling Program) have
drilled more than 2800 boreholes since 1968. The LILY database
(Laboratory Information management system with LithologY) aggregates
measurements from a subset of those expeditions \cite{lily_paper}. These
archives are sparse by construction. Different instruments sample at
different intervals, core recovery is incomplete, and a destructive
analysis such as crushing a sample for geochemistry precludes later
physical measurement of that material. Any given pair of measurements
therefore co-occurs on only a small fraction of the archive.
Measurements are typically paired with lithology, the rock type assigned
by a geologist on the basis of color, mineralogy, grain size and
texture. Downstream scientific analysis typically relies on the
lithology label, which is typically treated as the ground truth that a
model should be taught to reproduce
\cite{hall2016facies,hall2017distributed}. Here, however, we report the
opposite result.

Our result is based on a variational autoencoder \cite{kingma2013auto},
a neural network trained to compress each sample into a single vector.
Three properties of the archive shape the model. Measurements are often
missing, so during training we hide some inputs and replace them with
learned placeholder values, as in masked autoencoders
\cite{he2021masked}. The six inputs have different noise levels, so the
decoder learns a separate noise scale for each. The embedding should
also place similar rock close together, which the variational prior
encourages \cite{higgins2017betavae}. We compare the self-supervised
model on the same six inputs and the same 30 borehole folds with
principal component analysis, a semi-supervised variant, two
self-supervised tabular foundation models, TabNet \cite{arik2021tabnet}
and SCARF \cite{bahri2022scarf}, along with ablations that remove each
part of the model (Table~\ref{tab:main}). Our model separates lithology
better than all three.

We train using six properties that a geologist would use to classify a
rock: how dense is it, is it magnetic, is it radioactive, and what color
is it. Those are bulk density measured by gamma ray attenuation (GRA),
magnetic susceptibility (MS), natural gamma radiation (NGR), and red,
green and blue reflectance (RGB) from line scan imaging. They are the
measurable basis of lithological identity, which is the reason a
self-supervised model recovers lithology from them and, as
Section~\ref{sec:labels} shows, the reason the labels turn out to be
redundant.

\section{Data}
\label{sec:data}

\subsection{The LILY database}

The LILY database was built by \citeA{lily_paper} from the IODP
Laboratory Information Management System. It pairs each physical,
chemical and magnetic measurement with the lithology described at the
same depth in the core, covering more than 34 million measurements over
89 km of core from 42 expeditions drilled between 2009 and 2019. Each
lithology is recorded as a prefix, a principal name and a suffix, and
LILY adds simplified classes and a lithology type (sedimentary, igneous
or metamorphic) so that descriptions written by different shipboard
scientists can be compared. The paper also corrects whole-core gamma ray
attenuation bulk density against discrete moisture-and-density
measurements, and uses grain densities for each lithology to estimate
porosity. The database is public on Zenodo \cite{lily_zenodo}.

\subsection{Building the training set}

\begin{figure}
\centering
\includegraphics[width=\textwidth]{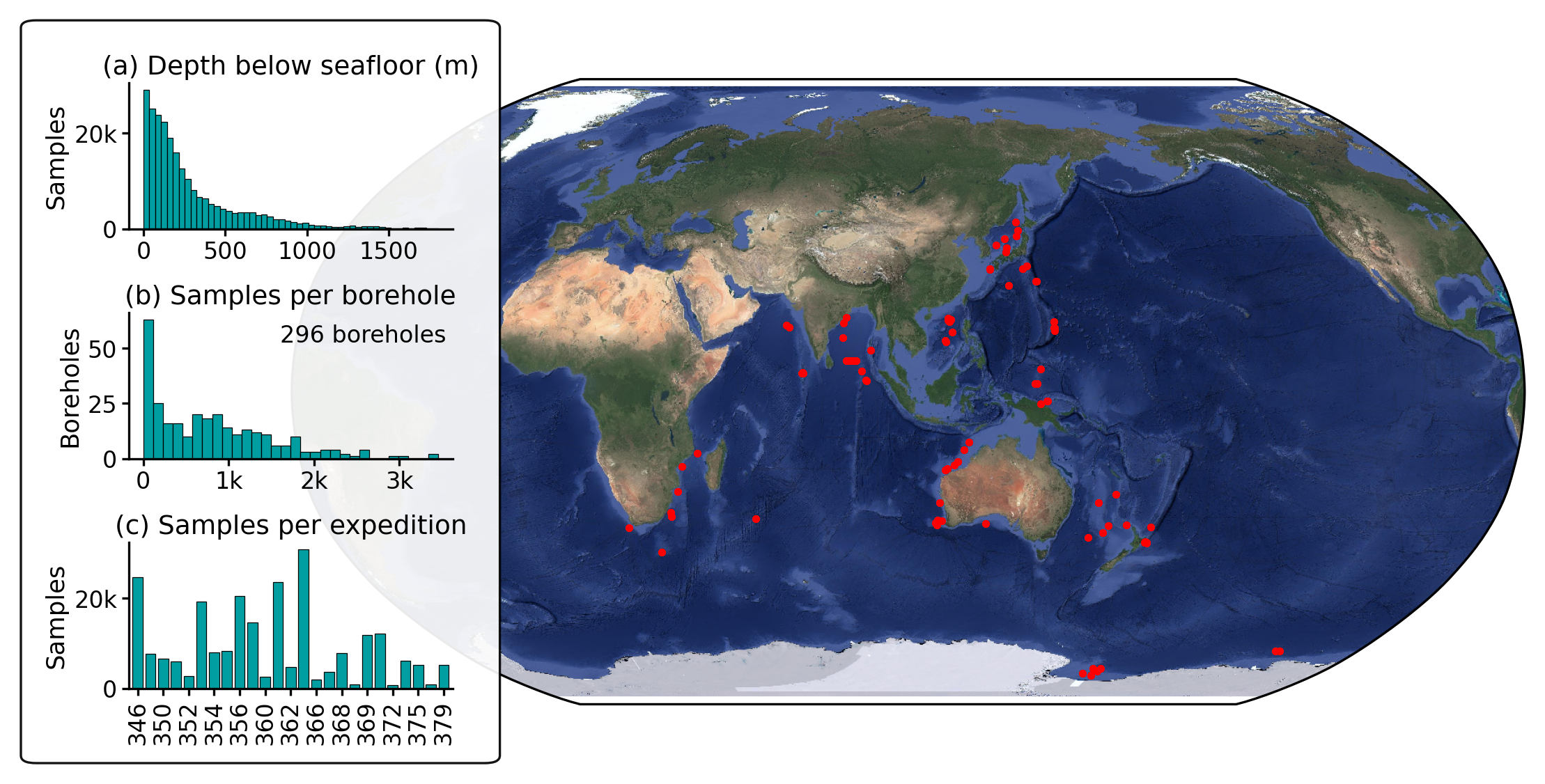}
\caption{Boreholes contributing to the modeled subset, with the depth
distribution, the samples per borehole and the samples per expedition.
The holes shown were drilled from several platforms.}
\label{fig:dataset}
\end{figure}

Measurements are binned into 20~cm depth intervals so that readings from
different instruments can be co-located (Figure~\ref{fig:dataset}). We
keep only bins in which all six properties are present. We then remove
rows whose values cannot be measurements, of which bulk density is the
clearest case: the release carries 74 negative values and 59 above 3.0~g/cm$^3$,
against a sediment median of 1.71 and a maximum plausible
value near 2.9 for dense basalt. Those 133 rows (0.056\% of the column)
have an effect out of all proportion to their number, because extreme
values dominate the sum of squares. Left in, they give two of thirty
folds a reconstruction $R^2$ of $-6\times10^{5}$ and $-82$, while the
median stays at 0.997. That leaves 236,963 observations across 296
boreholes from 25 expeditions, carrying 139 lithology classes. The
classes are heavily skewed, and two of them (nannofossil ooze and clay)
together account for over a third of the observations.

The subset is small relative to the full database precisely because the
six properties rarely co-occur in the same 20~cm bin. This also explains
why marine sediment dominates the subset. Whole-round logging assumes an
intact, full-diameter core, which soft sediment recovered by advanced
piston coring provides and fractured, low-recovery igneous or basement
rock recovered by rotary coring does not, so readings taken across voids
and rubble are invalid and we drop them. The dominance of sediment in
the subset is a consequence of that overlap.

\subsection{Held-out variables}
\label{sec:targets}

We hold out variables in LILY to measure zero-shot prediction from the
learned representation. We also remove some of the variables from LILY.
Variables were removed for one of two reasons. Some describe how the
core was drilled or measured and nothing about the geology, such as the
applied demagnetizing field, caliper separation, sample-preparation
masses, record identifiers, and magnetic declination and inclination,
which in unoriented core record how each piece sat in the barrel and the
overprint from the drill string. Others are close to linear
transformations of another variable, such as the moisture-and-density
family (MAD), CIELAB coordinates computed from tristimulus values, and
raw and drift-corrected magnetometer readings, so we keep one of each.

The variables we keep fall into two groups. The first measures a
training property with a different instrument, such as point magnetic
susceptibility and spectrophotometric reflectance (RSC), and recovering
these tests transfer across sensors. The second holds independent
physical properties the model never sees, including porosity, thermal
conductivity, discrete P-wave velocity and whole-core magnetization.

\begin{figure}
\centering
\includegraphics[width=\textwidth]{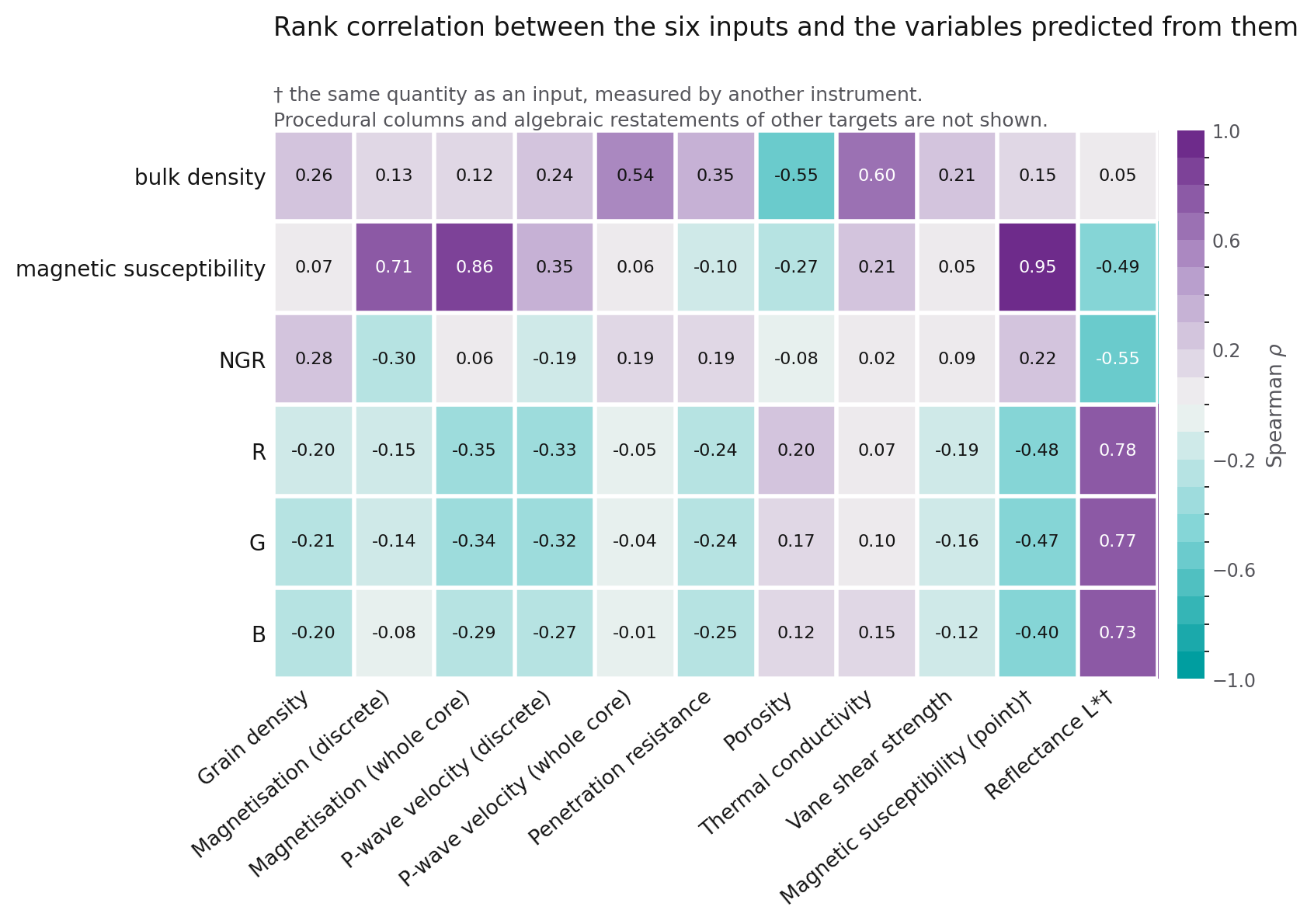}
\caption{Spearman rank correlation between each of the six inputs and each
reported held-out variable. Point magnetic susceptibility follows the
loop susceptibility input at $\rho = 0.95$ and reflectance follows red
at 0.78, so both are marked $\dagger$ as the same quantity as an input,
measured by another instrument, and not counted as new physics. Bulk
density carries porosity at $-0.55$ and thermal conductivity at 0.60.
Columns that are procedural or algebraic restatements of other columns
are not shown.}
\label{fig:correlations}
\end{figure}

Figure~\ref{fig:correlations} shows the rank correlation between the six
inputs and the variables we keep.

\section{Methods}
\label{sec:methods}

\subsection{Representation learning and variational autoencoders}

Representation learning trains a model to turn raw measurements into a
short list of numbers, an embedding, that keeps what matters about each
sample and drops the rest. A good embedding places similar samples close
together, so that simple methods applied to it, such as clustering or a
linear classifier, can answer questions the model was never trained on.
A geologist wants one description of a sample that can be reused for
lithology, porosity, thermal conductivity or any property measured
later, without training a new model each time. How well a model fits its
training data says little about whether its embedding is informative. A
model instead must be tested by asking the embedding to predict things
it was never shown, on boreholes it has never seen
(Section~\ref{sec:protocol}).

An autoencoder is a pair of neural networks trained together. The
encoder compresses each input into an embedding, and the decoder
reconstructs the input from that embedding. Because the embedding is
smaller than the input, the model can only reconstruct well by keeping
the structure that samples share. A variational autoencoder
\cite{kingma2013auto} extends this approach in two ways: (1) The encoder
outputs a probability distribution for each sample, a mean and a spread
in every embedding dimension, in place of a single point. (2) Training
penalizes those distributions for straying from a standard normal
distribution, which keeps the embedding space smooth so that nearby
points decode to similar measurements. The weight on that penalty,
$\beta$ \cite{higgins2017betavae}, sets the trade-off between
reconstruction accuracy and regularization toward the prior. While the
model itself needs no labels to train, we include lithology labels in a
semi-supervised variant described below, to test whether they change the
embedding.

\subsection{Architecture}

The encoder maps a masked input $\mathbf{x} \in \mathbb{R}^6$ through
two hidden layers (128 and 64 units, rectified linear activations) to
the mean $\boldsymbol{\mu}_z \in \mathbb{R}^{10}$ and log-variance $\log
\boldsymbol{\sigma}^2_z$ of the approximate posterior
$q(\mathbf{z}|\mathbf{x}) = \mathcal{N}(\boldsymbol{\mu}_z,
\mathrm{diag}(\boldsymbol{\sigma}^2_z))$. Each latent dimension is
sampled as $z_i = \mu_{z,i} + \sigma_{z,i}\epsilon_i$ with $\epsilon_i
\sim \mathcal{N}(0,1)$.

The decoder maps $\mathbf{z}$ through two hidden layers (64 and 128
units) to both a reconstructed mean $\hat{\boldsymbol{\mu}}_x \in
\mathbb{R}^6$ and a per-variable log-variance
$\log\hat{\boldsymbol{\sigma}}^2_x$, clamped to $[-6, 6]$ for numerical
stability. The six properties have very different noise scales. Magnetic
susceptibility spans thousands of instrument units while bulk density
varies by fractions of a g/cm$^3$. A decoder with one fixed variance is
dominated by whichever property has the largest residuals, so the
per-variable variance prevents a single noisy input from dominating the
geometry of the latent space.

Masking hides part of each input during training and asks the model to
reconstruct it from what remains. It thus forces the model to learn relationships among the variables. This is the idea behind masked pretraining in language
\cite{devlin2018bert} and images \cite{he2021masked}, and behind
self-supervised learning on tabular data \cite{yoon2020vime}. It suits
this archive in particular, where measurements are often missing,
because a model trained to fill in hidden values learns to work from
whichever measurements are present in a given sample.

During training we mask 10\% of input features at random, and replace
the masked values by learnable parameters $\mathbf{m} \in \mathbb{R}^6$
and not by zero. After standardization a zero is the population mean, so
the model would read it as a real measurement.

\subsection{Objective}

The self-supervised loss combines the heteroscedastic reconstruction
negative log-likelihood with a Kullback-Leibler (KL) term
\cite{kingma2013auto}:

\begin{equation}
\mathcal{L}_{\mathrm{self}} = \frac{1}{2}\sum_{n=1}^{6}
\left[\log\hat{\sigma}^2_{x,n} +
\frac{(x_n - \hat{\mu}_{x,n})^2}{\hat{\sigma}^2_{x,n}}\right]
+ \beta\, D_{\mathrm{KL}}\!\left[q(\mathbf{z}|\mathbf{x}) \,\|\,
p(\mathbf{z})\right],
\label{eq:loss-self}
\end{equation}

with $p(\mathbf{z}) = \mathcal{N}(\mathbf{0}, \mathbf{I})$.

The semi-supervised variant adds a linear classification head from
$\mathbf{z}$ to 139 lithology logits, weighted by $\alpha$:

\begin{equation}
\mathcal{L}_{\mathrm{semi}} = \mathcal{L}_{\mathrm{self}}
+ \alpha \sum_{c=1}^{139} -y_c \log \hat{y}_c.
\label{eq:loss-semi}
\end{equation}

We anneal $\beta$ linearly from $10^{-10}$ to $0.075$ over the first 50
epochs and hold it for 50 more
\cite{burgess2018understandingdisentanglingbetavae}; $\alpha$ anneals
from 0 to its final value over the same window. We optimize with Adam
\cite{kingma2015adam} (learning rate $10^{-3}$, weight decay $10^{-5}$,
batch size 256, gradient clipping at norm 1.0).

We do not correct the imbalance between lithologies by reweighting. The
lithology term in Equation~\ref{eq:loss-semi} is an unweighted
cross-entropy, and we report a macro-averaged classification metric, so
every lithology counts equally regardless of its prevalence. Lithologies
with fewer than two examples on a given side of a split cannot be scored
and are skipped; the number of lithologies scored is reported with each
result.

\subsection{Evaluation protocol}
\label{sec:protocol}

Two properties of the data constrain evaluation of the model. Boreholes
are geologically non-exchangeable. First, one borehole can carry
lithologies that appear nowhere else in the database. Second,
measurements within a borehole are strongly autocorrelated along depth,
therefore a random split over samples places near-duplicate rows in both
training and test data which causes the model to memorize the core
rather than to learn to generalize from it.

We therefore exclude random sample-level splitting and split by borehole
instead. Any single borehole split is biased by whichever boreholes
happen to be held out. Therefore every held-out number here is a median
with an interquartile range over 30 borehole-level folds, each holding
out 10\% of boreholes, with the model retrained from scratch on every
fold. The spread across folds measures that non-exchangeability and is
treated as a quantity of interest and not as noise. Blocked
cross-validation of this kind is standard where samples carry spatial
structure \cite{roberts2017cv,ploton2020spatial}. Every model in this
paper, including the baselines and the ablations, is evaluated on the
same folds, so rows of the main results table differ by architecture,
not by protocol.

We report two distinct quantities when we predict held-out variables.
\emph{Recoverability} fits the downstream regressor on the same rows it
scores, and asks whether a variable is present in the representation at
all. \emph{Transfer} retrains the encoder on a fold's training
boreholes, fits the regressor on those boreholes only, and scores on
boreholes that neither has seen, in order to determine how the relation
carries to new rocks. Each fold additionally records how far its
held-out boreholes sit from its training boreholes, as the Jaccard
overlap of their lithology sets \cite{jaccard1901etude} and as an energy
distance in the six input features \cite{szekely2013energy}. This is
what makes our protocol falsifiable, because holding out whole boreholes
invites the objection that the held-out rock is not drawn from the same
population as the training rock, so the resulting score would understate
what the model would achieve on a new borehole. Recording the distance
lets us test for this case (Section~\ref{sec:holdout-bias}).

We score reconstruction as per-variable $R^2$ on held-out boreholes, and
we cluster with a ten-component Gaussian mixture model on the
embeddings, scored by adjusted Rand index (ARI) against the lithology
labels. We classify with a per-class one-versus-rest support vector
machine on the embeddings, scored by macro-averaged area under the
receiver operating characteristic curve (AUC), and we predict the
held-out variables with CatBoost \cite{catboost} (500 trees of depth 6)
on the ten-dimensional embeddings.

\section{Results}
\label{sec:results}

\subsection{Baselines and ablations}
\label{sec:baselines}

\begin{table}
\caption{Every model on the same 30 borehole-level folds, reported as
median [interquartile range] over folds. Reconstruction $R^2$ and
clustering are measured on held-out boreholes. The AUC column is a
one-versus-rest linear support vector machine fitted on the held-out
embeddings, so it measures how
separable lithology is in the representation. PCA on six features is a rotation, so its
reconstruction is trivially exact. TabNet and SCARF do not reconstruct.}
\label{tab:main}
\centering
\footnotesize
\begin{tabular}{lccc}
\hline
Model & $R^2$ (recon) & ARI & AUC (probe) \\
\hline
\multicolumn{4}{l}{\textit{Full models}} \\
Self-supervised VAE & 0.999 [0.999, 1.000] & 0.237 [0.204, 0.283] & 0.958 [0.956, 0.965] \\
Semi-supervised VAE & 0.999 [0.999, 0.999] & 0.247 [0.214, 0.290] & 0.959 [0.955, 0.964] \\
\hline
\multicolumn{4}{l}{\textit{Ablations, one component changed}} \\
Fixed-variance decoder & 0.976 [0.972, 0.982] & 0.224 [0.206, 0.267] & 0.939 [0.933, 0.949] \\
Zero masking & 0.999 [0.999, 1.000] & 0.237 [0.203, 0.280] & 0.960 [0.957, 0.963] \\
No masking & 1.000 [1.000, 1.000] & 0.248 [0.224, 0.277] & 0.950 [0.946, 0.956] \\
Fixed $\beta$ & 0.999 [0.999, 0.999] & 0.254 [0.207, 0.306] & 0.958 [0.955, 0.962] \\
\hline
\multicolumn{4}{l}{\textit{Baselines, no labels}} \\
PCA & 1.000 [1.000, 1.000] & 0.203 [0.168, 0.254] & 0.939 [0.936, 0.947] \\
TabNet & -- & 0.183 [0.157, 0.224] & 0.951 [0.947, 0.957] \\
SCARF & -- & 0.167 [0.138, 0.199] & 0.953 [0.943, 0.958] \\
\hline
\end{tabular}
\end{table}

Because every model in Table~\ref{tab:main} is scored on the same 30
borehole folds, we compare models pairwise with a Wilcoxon signed-rank
test on the per-fold differences and call a difference resolvable at $p
< 0.05$. Measured this way, the fixed-$\beta$ and no-masking ablations
are indistinguishable from the full model on clustering ($p = 0.08$ and
$p = 0.10$).

Three differences hold up. Removing the heteroscedastic decoder makes a
small difference that the paired test resolves. Reconstruction falls
from 0.999 to 0.976 and classification from 0.958 to 0.939
($\Delta\mathrm{ARI} = 0.016$, $p = 0.016$; $\Delta\mathrm{AUC} =
0.019$, $p < 10^{-4}$), which is what we expect if a single noise scale
is set by whichever input has the largest residuals. The learned
representation separates lithology better than principal components on
the same six inputs ($\Delta\mathrm{ARI} = 0.030$, $p = 0.0007$), a
difference that the overlapping interquartile ranges in
Table~\ref{tab:main} would hide. And it separates lithology better than
either self-supervised tabular baseline (TabNet \cite{arik2021tabnet}
and SCARF \cite{bahri2022scarf}), by 0.065 and 0.059 adjusted Rand
index, both resolvable at $p < 10^{-4}$. The $\beta$ schedule and the
masking strategy shift classification by 0.002 and 0.009 respectively;
both clear the threshold on 30 paired folds and neither is large enough
to matter.

Every model capable of reconstruction achieves an $R^2$ of 0.999 or above, and principal components on six features reach exactly 1.000 (six components of a
six-dimensional input are a rotation). Because even a trivial baseline is perfect, we thus conclude that regression accuracy is not informative for this task.

We thus evaluate the intrinsic difficulty of the task by directly
training a classifier on the six raw inputs with lithology labels. The
classifier reaches 0.900 transfer AUC, against 0.838 from a probe on the
embedding and 0.820 from a probe on the raw inputs. Predicting held-out
variables, the raw inputs do marginally better than the embedding (0.225
against 0.201), and so do principal components on the same six inputs
(0.218). A kNN predictor on the raw inputs is the only model that falls
behind (0.100). The encoder maps six standardized inputs to ten
dimensions and so cannot add information. Classification and
reconstruction are therefore not the hard part of the task. The hard
part is building a latent space whose structure describes the physical
properties behind the measurements, and a model that scores well has not
shown that it scores well for the right reason.

We use the variational autoencoder for the rest of the paper because it
gives a continuous probabilistic latent space to examine, with no
assumption that the measurements relate linearly. TabNet and SCARF do not create latent spaces, and principal components form latent space only under a linear assumption.

\subsection{Supervision by lithology label adds nothing}
\label{sec:labels}

We find that the two full models are indistinguishable. Over 30 borehole
folds the self-supervised model reaches an adjusted Rand index of 0.237
[0.204, 0.283] against the lithology labels, and the variant trained
with those labels reaches 0.247 [0.214, 0.290]. Classification is 0.958
[0.956, 0.965] against 0.959 [0.955, 0.964], and reconstruction is 0.999
for both (Table~\ref{tab:main}). We compared the two models fold by fold
with a Wilcoxon signed-rank test, and none of the differences is
significant (clustering $p = 0.38$, classification $p = 0.42$,
reconstruction $p = 0.79$). The same test detects a difference of 0.030
in adjusted Rand index between the self-supervised model and principal
components, so it is sensitive enough to find a difference of that size.
We find that adding the lithology labels does not change the result.

Shipboard lithology is assigned by eye from color, grain size, texture
and induration, and several of our inputs reflect these properties.
Reflectance in three channels is color, and bulk density responds to
induration and porosity. Natural gamma radiation indexes clay content
through its potassium, uranium and thorium (K, U and Th)
\cite{vleeschouwer2017quantifying}, and magnetic susceptibility indexes
the ferrimagnetic fraction \cite{ellwood1980induced}. A model with
access to these six features thus has access to most of the information
available to a geologist, and so the label itself provides little
additional information.

A macro average does not capture differential accuracy across classes, particularly when there is strong class imbalance, so we also compare the two by lithology, on
the 89 classes both could score on the same held-out folds. If the
labels carried anything the measurements lack, it should appear on the
classes the measurements separate worst. It does not. The
semi-supervised model is better by more than 0.01 on 27 classes, worse
on 28, and within 0.01 on 34, with a median difference of $-0.002$
(Wilcoxon $p = 0.41$), and the rarer and commoner halves of the class
distribution are indistinguishable ($-0.0023$ against $-0.0020$).

\begin{figure}
\centering
\includegraphics[width=\textwidth]{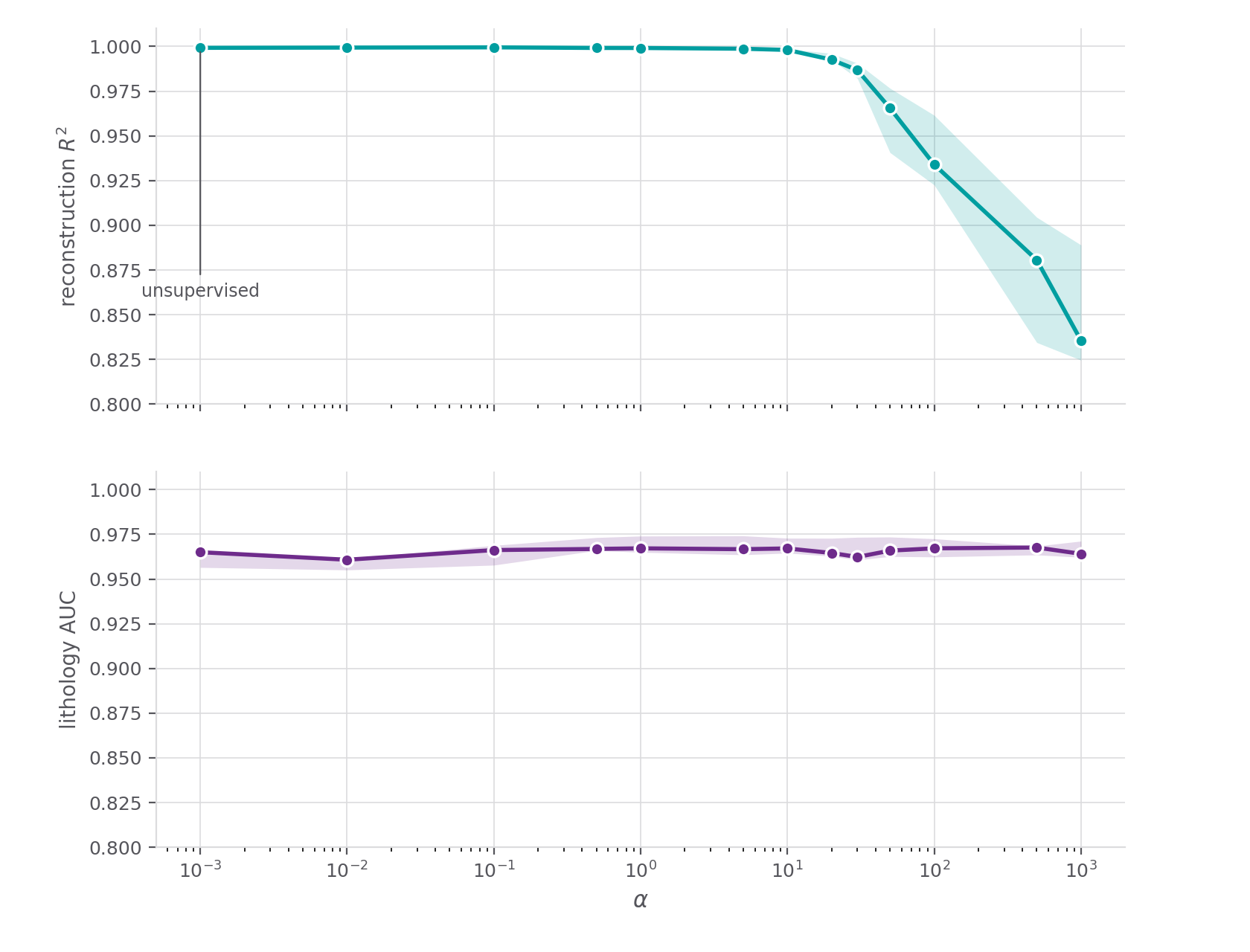}
\caption{Supervision weight against reconstruction and classification, median
over five borehole folds with the interquartile range shaded. The
leftmost point is the fully self-supervised model, drawn at a tenth of
the smallest non-zero weight so that it can be placed on a logarithmic
axis. Reconstruction is unaffected until $\alpha$ reaches about 10 and
then falls steadily; classification is flat across the whole range.}
\label{fig:alpha}
\end{figure}

Supervision does not increase reconstruction and begins to actively harm
it as it becomes more important (Figure~\ref{fig:alpha}). Reconstruction
falls monotonically from 0.999 to 0.836 as $\alpha$ rises from zero to
1000 (Spearman rank correlation $\rho = -0.98$ against $\log\alpha$),
while lithology classification stays between 0.961 and 0.968 throughout
($\rho = +0.20$, $p = 0.51$) and clustering shows no trend ($\rho =
+0.09$, $p = 0.78$). Even when the classification term is weighted a
thousand times the reconstruction term, the model classifies lithology
no better than when it has never seen a label, and the only thing the
weight buys is a worse account of the measurements. With five folds a
signed-rank test cannot resolve any individual $\alpha$, since its
smallest attainable $p$ is 0.0625; the trend across all thirteen carries
the result.

\subsection{Zero-shot prediction}
\label{sec:zeroshot}

\begin{figure}
\centering
\includegraphics[width=\textwidth]{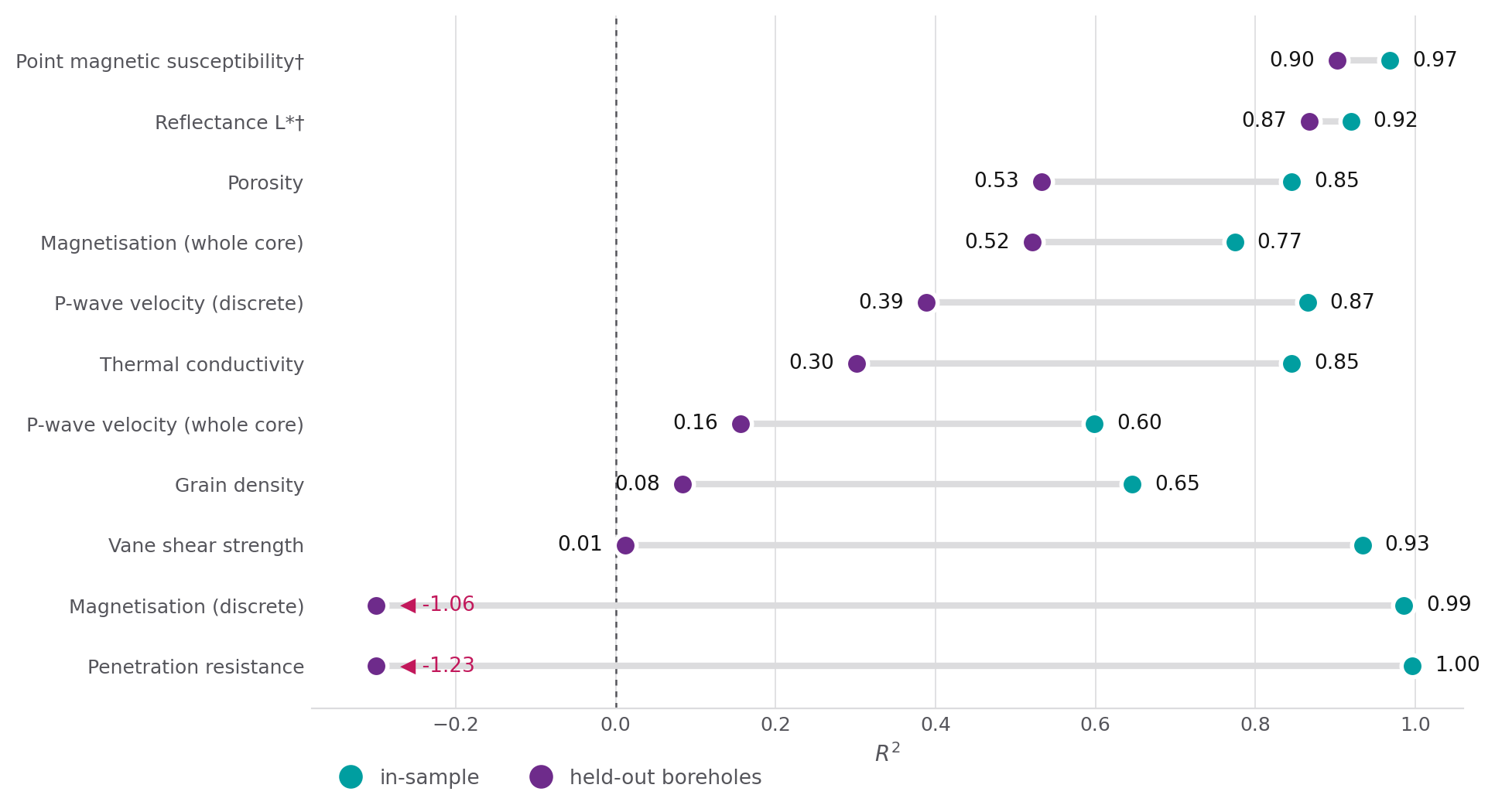}
\caption{Each held-out variable predicted from the ten-dimensional
self-supervised embedding, by gradient-boosted regression trees
(CatBoost, 500 trees of depth 6), scored two ways: on the boreholes the
regressor was fitted on, and on boreholes that neither it nor the
encoder had seen. Median over 30 borehole folds. Every held-out variable
is shown, including the ones that fail. Three bands are visible.
Cross-sensor targets, marked $\dagger$, barely move. Physical properties
lose a third to a half. The outright failures were near-perfect before
the holdout. $\blacktriangleleft$ marks a value below the axis, printed
beside it.}
\label{fig:train-vs-heldout}
\end{figure}

Figure~\ref{fig:train-vs-heldout} reports each held-out variable twice.
It is scored on the boreholes the downstream regressor was fitted on,
and scored on boreholes that neither it nor the encoder had seen. The
distance between the two measures the ability of the model to generalize
across boreholes on unseen variables.

Porosity reaches 0.53, whole-core magnetization 0.52, discrete P-wave
velocity 0.39 and thermal conductivity 0.30, on boreholes the model has
never seen. These are not arbitrary successes, and the couplings that
produce them have been previously described. Bulk density fixes porosity
through the mass balance between grain, pore fluid and bulk volume in
saturated sediment \cite{blum2003}. Thermal conductivity follows from
porosity and the mineral framework, as a mixture of the conductivity of
the solid and of the pore water
\cite{goto2009relations,brigaud1989mineralogy}, and does so nonlinearly
\cite{ghanbarian2016thermal}. Compressional velocity in unconsolidated
marine sediment follows porosity and grain size
\cite{hamilton1982sound,li2019measurement}. Remanent magnetization and
magnetic susceptibility both scale with the concentration of
ferrimagnetic minerals \cite{ellwood1980induced}. The model has not
discovered these relations; it has exploited them, which is the reason
to expect the same approach to work on other archives built from the
same measurements.

Magnetic susceptibility and color also perform well, but they do not
represent true transfer. Point magnetic susceptibility reaches $R^2 =
0.90$ and spectrophotometric reflectance 0.87, and both barely move when
we impose the holdout, falling 0.07 and 0.05 respectively. Both are
training inputs measured by a different instrument. The representation
is thus not tied to the sensor it was built from, and we report these
two separately from the independent targets.

Two mechanical properties fail. Vane shear strength collapses from 0.93
on the boreholes the regressor saw, to 0.01 on those it did not, and
penetration resistance falls from 1.00 to $-1.23$. Both are mechanical
properties of the sediment fabric, set by consolidation history,
cementation and disturbance during coring, and none of the six inputs
measures these directly. Their near-perfect in-sample scores are a result of the regressor scoring itself on the core it had been fitted on, rather than predicting a hard property well.

\subsection{Depth does not transfer between boreholes}
\label{sec:depth}

Sediment accumulates by superposition and compacts under its own weight.
Depth ought therefore to be a strong control on almost any physical
property, and it is possible that the model has learned burial and
nothing else. We test this by predicting each held-out variable from
depth alone, on the same 30 borehole folds used for the embedding. Depth
alone reaches a median $R^2$ of $-0.017$ across the nine independent
targets, against 0.201 for the embedding. It carries some signal for
porosity (0.30) and discrete P-wave velocity (0.31), the two properties
most directly set by compaction, but the embedding does better on both
(0.53 and 0.39). The embedding does better on eight of the nine targets;
the exception is discrete magnetization, which neither predicts. Depth
does not transfer across sites. The sedimentation rate, the sediment
source, the geothermal gradient and the burial history all differ across
boreholes, so the relation between depth and any property differs from
hole to hole. Around half the variance in these properties (median 52\%)
lies between boreholes, not within them. Depth fails to transfer whether
we use an absolute depth (i.e., normalized from the top of the hole or
sea level) or a relative depth (i.e., position within each hole's own
cored interval, measured from the top).

\section{Discussion}
\label{sec:discussion}

\subsection{Expert labels and self-supervision}

Whether models should learn from expert labels or from the structure of
unlabeled data is an open question in machine learning. Labels are
expensive, and annotators often disagree in ways that a single, correct
label hides \cite{plank2022problem,grossmann2022beyond}. Self-supervised
pretraining, which learns from unlabeled data alone, now matches or
beats supervised training in language \cite{devlin2018bert} and images
\cite{he2021masked}. In medical imaging it generalizes to new hospitals
with far fewer labeled examples \cite{azizi2023robust}, and in biology a
model trained only on protein sequences learns structural and functional
properties that were never labeled \cite{rives2021biological}.

Lithology is a case where the answer is unusually clear. The
information that defines a lithology is already in the measurements
(Section~\ref{sec:labels}). The exception is fossil content. Telling nannofossil ooze from
foraminiferal ooze needs a microscope, and none of the six instruments
records it, so for those distinctions the label carries information the
measurements do not.

The categories do cost us something, because lithology is a partition
imposed on quantities that vary continuously. A sediment does not become
clay at some threshold of grain size, it grades. Assigning a name
compresses a position in a continuous space to a symbol, and it is lossy
in a specific way, because it discards exactly the gradational detail
that records how the sediment accumulated. A model trained on the
measurements keeps that information. A model trained on the labels
cannot recover it. We do not find that geologists are wrong. We find
that categorization, which is how the field has stored its observations
for two centuries, is a lower-fidelity record of process than the
instruments already sitting on the same core.

\subsection{Mixed-determined problems in Earth science}

The difficulty this work addresses is not specific to drill core
\cite{bergen2019machine,reichstein2019deep}. Earth science data are
heterogeneous by construction: different instruments sample at different
resolutions, coverage is uneven and set largely by where it was
convenient to measure, and the quantities of interest are usually not
the ones recorded. The resulting inverse problems are mixed-determined
\cite{menke2012geophysical}, overdetermined in the few places with dense
sampling and underdetermined everywhere else, with hidden variables and
scaling laws that are rarely known in advance.

A learned representation over whatever was measured is one way into that
class of problem, and sediment is a convenient place to test it because
the couplings are already understood well enough to check the answer.
The same structure turns up elsewhere. Seismic hazard sets dense
instrumental records against sparse paleoseismic constraints. In
critical mineral and hydrogen exploration a handful of assayed holes has
to inform a whole region. In subsurface storage the property that
matters is permeability which is not measured. Groundwater and glacier
instability repeat the same mismatch between what is measured and what
is wanted. What we can offer those problems is not the architecture,
which is ordinary, it is the evaluation. Hold out the unit of geological
independence, check whether the signal outlasts the strongest
confounding coordinate, and verify that the variables being predicted
are not restatements of the inputs.

\subsection{Depth and compaction}

Section~\ref{sec:depth} shows that depth alone cannot predict properties of unseen boreholes. Within a single hole, depth orders the
measurements closely and is easily learned by a model, so any model fitted
to borehole data without blocking by borehole performs well by learning to predict depth. When evaluated on rows drawn from
the same holes, such a model will produce high scores. However, a model dependent on predicting depth will fail on the
next hole.

Regressing depth out of each target separates depth-driven from
rock-driven prediction. On held-out boreholes over the same 30 folds,
the embedding recovers the residualized targets at 0.253 for thermal
conductivity, 0.432 for whole-core magnetization and 0.169 for discrete
P-wave velocity, against 0.301, 0.521 and 0.388 before depth is removed.
The two scores are not on one scale, since removing depth also removes
variance. Velocity is largely compaction and should be described as
such; conductivity and magnetization are not.

\subsection{Beyond marine sediment}

Whether a representation trained on marine sediment carries to anything
else is untested here and is the obvious next question. The six
measurements are not specific to ocean drilling. Density, magnetic
susceptibility, natural gamma radiation and color are logged on
continental scientific cores, on mineral exploration core, and in
petroleum boreholes, usually with the same or similar instruments. The
couplings invoked in Section~\ref{sec:zeroshot} hold for saturated
porous media anywhere, so there is reason to expect them to hold there
too. What will not carry is the range. This subset spans soft sediment
in the upper part of the section, and the relations between porosity,
velocity and conductivity change character once cementation and
lithification set in. A model fitted here will thus fail on crystalline
basement, and testing exactly where it begins to fail would be more
informative than extending it.

\subsection{Data infrastructure is the bottleneck}

The limiting factor in this work is primarily data. Only 236,963
intervals in the database of 34 million measurements have all six
properties, which can bias reported metrics. However, measuring these
six properties is cheap and non-destructive, and so an archived core in
the IODP repositories (MARUM, College Station, and Kochi) could be
rescanned to fill the gaps, and ingesting the rest of the IODP holdings
rather than the subset LILY happened to cover would multiply the
available boreholes several times over. Validity ranges on stored
quantities would have caught the corrupt values when they were
deposited, not a decade later. The field has spent forty years
generating these measurements and comparatively little making them
jointly usable, and it is that second effort, not a larger model, that
determines what can be learned.

\section{Conclusions}
\label{sec:conclusions}

Six measurements that are logged on ocean drilling cores as a matter of
routine carry enough information to recover the lithological
classifications a geologist assigns. Training a model that includes
those labels degrades the model's account of those geophysical
measurements. Clustering, classification and reconstruction are
indistinguishable between the two variants across 30 borehole folds, and
no individual lithology benefits. Raising the weight on the labels
through six orders of magnitude degrades the model's account of the
measurements and leaves classification untouched. The labels compress
the measurements and add nothing to them.

The same representation recovers physical properties that were never
part of training, on boreholes the model has not seen: porosity at $R^2
= 0.53$, whole-core magnetization at 0.52, discrete P-wave velocity at
0.39, thermal conductivity at 0.30. Each rests on a coupling that was in
the literature before this model existed, and the properties with no
such coupling to the six inputs fail completely.

Depth does not carry from one hole to the next. On boreholes it has not
seen, depth alone reaches a predictive median $R^2$ of $-0.02$ across
the independent targets, against 0.20 for the embedding.

We would emphasize the evaluation over the architecture. The model here
is ordinary, and measured on one protocol it is not distinguishable from
several simpler variants of itself. What changed our conclusions was
holding out the unit of geological independence instead of random rows,
checking that predicted variables were not restatements of the inputs,
removing values that cannot be measurements, and residualizing the
strongest confounding coordinate before claiming a relation. For
archives of this kind, that discipline is worth more than a larger
model.

\appendix

\section{Transfer is stable across held-out novelty}
\label{sec:holdout-bias}

To measure how far a fold's held-out boreholes sit from its training
boreholes we use the energy distance \cite{szekely2013energy}, a
standard test of whether two samples come from the same distribution.
For training rows $X$ and held-out rows $Y$ in the six standardized inputs
it is

\begin{equation}
\mathcal{E}(X, Y) = 2\,\mathbb{E}\lVert X - Y \rVert - \mathbb{E}\lVert X - X' \rVert - \mathbb{E}\lVert Y - Y' \rVert
\end{equation}

where $X'$ and $Y'$ are independent copies drawn from the same sets and the
norm is Euclidean. It is zero when the two sets share a distribution and
grows as they separate, with no assumption about the shape of either
distribution. The pairwise terms grow with the square of the number of
rows, so each side is subsampled to 2000 rows with a fixed seed per
fold. The value has no natural scale, so we calibrated it by shifting
one standardized input in a random tenth of the rows. The distance grows
roughly as the square of the shift, and the median fold value of 0.066
matches a shift of about 0.45 standard deviations.

Splitting by borehole could in principle depress every transfer score on
its own, by asking the model about rock that no training borehole
contains. Across the 30 folds the lithology overlap between training and
held-out sets has a median of 0.432 and ranges from 0.345 to 0.518, and
the energy distance in the six input features has a median of 0.066. If
the holdout were driving the result, folds with more novel rock would
score lower.

\begin{figure}
\centering
\includegraphics[width=\textwidth]{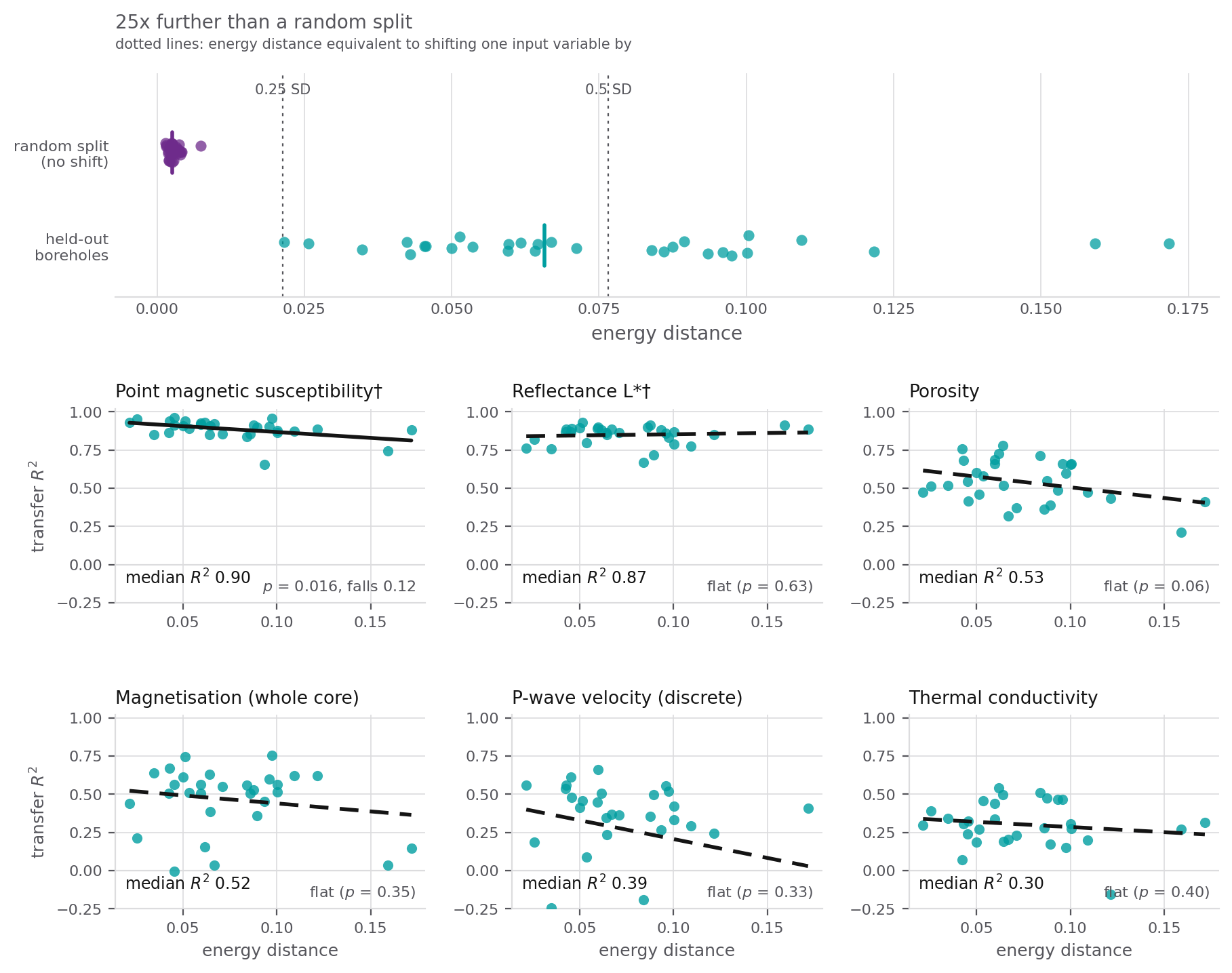}
\caption{Transfer against how far each fold's held-out boreholes sit from its
training boreholes, measured as an energy distance in the six
standardized inputs. The upper panel establishes that the held-out sets
really are out of distribution: they sit about 25 times further from the
training rows than a random split of the same data does, roughly a
0.45~standard-deviation displacement of one input. Of the six properties
that transfer, only point magnetic susceptibility declines
distinguishably from flat ($p = 0.016$), falling 0.12 from a median of
0.90. Fits that cannot be separated from flat are drawn dashed.}
\label{fig:holdout-novelty}
\end{figure}

They do not (Figure~\ref{fig:holdout-novelty}): transfer $R^2$ is
uncorrelated with both distance measures, at Spearman $\rho = -0.05$ ($p
= 0.78$) against lithology overlap and $\rho = +0.02$ ($p = 0.90$)
against energy distance. Extrapolating the linear fits to a held-out set
indistinguishable from training gives $R^2$ between 0.07 and 0.11,
against an observed median of 0.10. Splitting the folds at the median
overlap gives 0.120 for the more familiar half and 0.067 for the less
familiar half, a difference in the expected direction that this many
folds cannot separate from noise.

A direct comparison sharpens the question rather than settling it. For
porosity, a random split over rows scores $R^2 = 0.740$ against 0.531
for the borehole split, and for thermal conductivity the two are 0.474
and 0.291 (medians over the same 30 folds, with the encoder held fixed
so that only the regressor's split changes). The random split scores
higher for every target we tested. That is what
Section~\ref{sec:protocol} predicts, and it is not evidence that the
borehole holdout is pessimistic. Measurements sampled every 20~cm along
a core put near-duplicates of each held-out row into training, so a
random split scores the regressor partly on rows it has already seen,
which is the leak the borehole protocol exists to close.
What the comparison does give us is the size of that leak, close to 0.2
in $R^2$ for porosity, and it leaves the novelty analysis above to carry
the claim that the holdout itself is not what lowers the scores.

We should state one limitation of this test plainly. Lithology overlap
spans only 0.345 to 0.518 across the folds, so every fold holds out
substantial rock that training never saw, and the extrapolation to
complete overlap lies outside the range the data covers. The measurement
rules out a large pessimistic bias in the borehole holdout, and it
cannot rule out a small one.

\acknowledgments
This project received funding from the Norwegian Research Council
(SerpRateAI, grant no. 334395). TWB was partially supported by NSF EAR
2121666.

\section*{Open Research}

The LILY database is publicly available on Zenodo
\cite{lily_zenodo}. All code used to produce the results and figures in
this paper, including the preprocessing, the model, the evaluation
protocol described in Section~\ref{sec:protocol}, and the curation of
held-out variables described in Section~\ref{sec:targets}, is available
at \url{https://github.com/mnky9800n/hmvae-geo} and archived on
Zenodo \cite{hmvae_geo_zenodo}.

\bibliography{references}

\end{document}